\documentclass{PoS}

\usepackage{amsmath}
\usepackage{amssymb}
\usepackage{amsfonts}
\usepackage{graphics}
\usepackage{graphicx}
\usepackage{amscd}
\usepackage{amsfonts}
\usepackage{epsf}
\usepackage{epsfig}
\usepackage{color}
\usepackage{bm}

\def\be#1{\begin{equation}#1\end{equation}} 

\def\beqnn#1{\begin{eqnarray}#1\end{eqnarray}}
\def\exp#1{\mathrm{e}^{#1}} 
\def\integral#1#2{\int^{#2}_{#1}} 

\title{Test of factorization for the long-distance effects from charmonium on $B\to K\ell^+\ell^-$}

\ShortTitle{Factorization for $B\to K\ell^+\ell^-$}

\author{\speaker{Katsumasa Nakayama}$^{a, b}$, and Shoji Hashimoto$^{b,c}$ (JLQCD collaboration)\\
         $^a$ Department of Physics, Nagoya University, Nagoya, 464-8602, Japan\\
         $^b$ KEK Theory Center, High Energy Accelerator Research Organization (KEK), Tsukuba 305-0801, Japan\\
         $^c$ School of High Energy Accelerator Science, The Graduate University for Advanced Studies (Sokendai),Tsukuba 305-0801, Japan\\
        E-mail: \email{katumasa@post.kek.jp}}

\abstract{
We report on a calculation of the charmonium contribution to the decay $B \rightarrow  K\ell^+\ell^-$ using lattice simulations with 2+1 flavors of Mobius domain wall fermions.
We focus on the region of $q^2$ below the $J/\psi$ resonance and test the factorization approximation to estimate the amplitude.
We show a possible discrepancy between the lattice calculation and the factorization method for the non-factorizable contribution.
}

\FullConference{The 36th Annual International Symposium on Lattice Field Theory - LATTICE2018\\
        22-28 July, 2018\\
        Michigan State University, East Lansing, Michigan, USA.}

\begin{document}

\section{Introduction}

The rare decays $B\to K^{(*)}\ell^+\ell^-$ provide a unique probe of
new physics since the Standard Model contribution is suppressed by the
GIM mechanism. In fact, there are some hints of deviation from the Standard 
Model in the experimental data mainly from LHCb \cite{Aaij:2013pta, Aaij:2016cbx}.

Theoretically, an important question remains in the Standard Model
calculation of the corresponding amplitudes. Namely, the same
final state can be created by an intermediate charmonium state
decaying to $\ell^+\ell^-$, {\it i.e.} a decay chain $B\to K^{(*)}\psi \to
K^{(*)}\ell^+\ell^-$. The experimental analysis, therefore, treats the region of $q^2$, an invariant mass squared of the final lepton pair, away from the charmonium resonances.
It could still be a problem because the charmonium resonance contributions are so
large and even a small tail contribution can give a significant
effect to (off-resonance) $K^{(*)}\ell^+\ell^-$.
In other words, the long-distance effect from the charmonium
resonances have to be controlled.

Since the theoretical treatment of the multibody hadronic intermediate
state $K^{(*)}\psi$ is highly non-trivial, the processes are estimated
mainly using the factorization approximation where the two-body decay amplitude for $B\to K^{(*)}\psi$ is replaced by a product of a
semi-leptonic type amplitude $B\to K^{(*)}$ and an amplitude to produce $\psi$ from the vacuum. The correction to this rather naive
approximation has been discussed in the literature and no firm
estimate has been achieved so far \cite{Neubert:1997uc,Beneke:2001at,Lyon:2014hpa,Du:2015tda}.

In this work, we use lattice calculation to investigate the validity of the factorization approximation for $B\to K\ell^+\ell^-$.
We calculate a four-point function corresponding to the process
$B\to K\psi$ through the weak Hamiltonian $H_\mathrm{eff}$ as well as the amplitude for its factorization approximation obtained from two-point and three-point functions. We then explicitly test the factorization ansatz albeit at a slightly different kinematical setup
from the physical decay.

\section{Amplitude for $B\to K \ell^+\ell^-$}
We first describe the lattice computation of the decay amplitudes.
The formalism is along the same line of the study of $K\to \pi\ell^+\ell^-$ \cite{Christ:2015aha,Christ:2016eae}.
We consider a kinematical situation such that the formalism is made simple as discussed below.

We focus on the $B\to K\ell^+\ell^-$ through the charmonium resonances which could induce a significant effect on the amplitude.
The corresponding part of the effective weak Hamiltonian $H_\mathrm{eff}$ is written as
\be{
H_\mathrm{eff}
=
\frac{G_F}{\sqrt{2}}
V_{cs} ^*V_{cb}
\left(
C_1O_1 ^c
+
C_2O_2 ^c
\right),
}
where the Fermi constant $G_F$, CKM matrix elements $V_{cb}$ and $V_{cs}$, and the Wilson coefficients $C_i$ are introduced. 

The four-fermion operators $O_i ^c$ to produce a $c\overline{c}$ pair are represented as
\beqnn{
O_1 ^c
&=&
(\overline{s}_i\gamma_\mu P_-c_j)
(\overline{c}_j\gamma_\mu P_-b_i),\nonumber\\
O_2 ^c
&=&
(\overline{s}_i\gamma_\mu P_-c_i)
(\overline{c}_j\gamma_\mu P_-b_j),
}
where the indices $i$ and $j$ specify the color contraction, and $P_- \equiv \frac{1-\gamma_5}{2}$ is the projection operator.
Using the effective Hamiltonian $H_\mathrm{eff}$, we define the $B\to K\ell^+\ell^-$ decay amplitude with a four-momentum $q\equiv k-p$,
\be{
A(q^2)=\int\mathrm{d}^4x\ 
\exp{iqx}
\langle
K(\bm{p})|
T\left[
J_\mu(0)
H_\mathrm{eff}(x)
\right]
|B^j(\bm{k})
\rangle.
}
The vector current $J_\mu=\overline{c}\gamma_\mu c$ annihilates the $c\overline{c}$ pair to produce the lepton pair in the final state.

In order to obtain this amplitude, we calculate the four-point correlator $\Gamma_\mu ^{(4)}
(t_H,t_J,\bm{k},\bm{p})$ on the lattice,
\be{
\Gamma_\mu ^{(4)}
(t_H,t_J,\bm{k},\bm{p})
=
\integral{}{}\mathrm{d}^3\bm{x}
\integral{}{}\mathrm{d}^3\bm{y}
\ \exp{-i\bm{q}\cdot\bm{x}}
\langle
K(t_K,\bm{p})
|T\left[
J_\mu(t_J,\bm{x})
H_\mathrm{eff}(t_H,\bm{y})
\right]|B(0,\bm{k})
\rangle,
}
with Euclidean time-ordering $0<t_J,t_H<t_K$.
We then consider a time integration of the four-point correlator,
\be{
I_\mu
(T_a,T_b,\bm{k},\bm{p})
=
\ \exp{-\left[E_K(\bm{p}) - E_B(\bm{k})\right]t_J}
\integral{t_J - T_a}{t_J + T_b}
\mathrm{d}t_H
\Gamma_{\mu} ^{(4)}
(t_H,t_J,\bm{k},\bm{p}).
}
There are the contributions from various intermidiate states to this integral.
\beqnn{
I_\mu
&=&
-\integral{0}{\infty}\mathrm{d}E
\frac{\rho_1(E)}{2E}
\frac{
\langle K(\bm{p})|J_\mu(0)|E(\bm{k})\rangle
\langle E(\bm{k})|H_\mathrm{eff}(0)|B(\bm{k})\rangle
}
{
E_B(\bm{k})-E
}\left(
1 - \exp{(E_B(\bm{k})-E)T_a}
\right)
\nonumber\\
&&+
\integral{0}{\infty}\mathrm{d}E
\frac{\rho_2(E)}{2E}
\frac{
\langle K(\bm{p})|H_\mathrm{eff}(0)|E(\bm{p})\rangle
\langle E(\bm{p})|J_\mu(0)|B(\bm{k})\rangle
}
{
E-E_K(\bm{p})
}\left(
1 - \exp{-(E-E_K(\bm{p}))T_b}
\right).
}
Here, we use spectral densities $\rho_1(E)$ and $\rho_2(E)$ for the states 
with and without strangeness, respectively.
We then define the amplitude at $T_{a,b}\to \infty$ limit of $I_\mu$,
\be{
A(q^2)
=
-i\lim_{T_{a,b}\to \infty}
I(T_a,T_b,\bm{k},\bm{p}).
}

The integral suffers from a divergence, as in $K\to\pi \ell^+\ell^-$
\cite{
Christ:2015aha,
Christ:2016eae}.
An artificial divergence due to the term $\exp{(E_B(p) - E)T_a}$ in the limit of $T_a \to \infty$ has to be subtracted to obtain the physical amplitude.
In this work, instead of subtracting the divergent term, we restrict ourselves in a kinematical region where the divergence does not show up.
Namely, we take the $b$ quark mass such that the intermediate energy $E$ is larger than the $B$ meson mass, $E_B <  E$.
On the other hand, the intermediate energy is bounded by the ground state energies of the charmonium and $K$ meson, $E_{J/\psi} + E_K \leq E$.
We therefore require the condition $E_B < E_{J/\psi} + E_K$.
It means that we set the $b$-quark mass smaller than its physical value so that this condition is satisfied.
%
With this unphysical set-up, we can extract the decay amplitude from the four-point correlators.

In this work, we test the factorization approximation as a first step, leaving the calculation of the amplitude $I_\mu$ for future studies.


\section{Factorization of the four-point function}

\begin{figure}[tbp]
\begin{center}
  \includegraphics[width=10cm, angle=0]{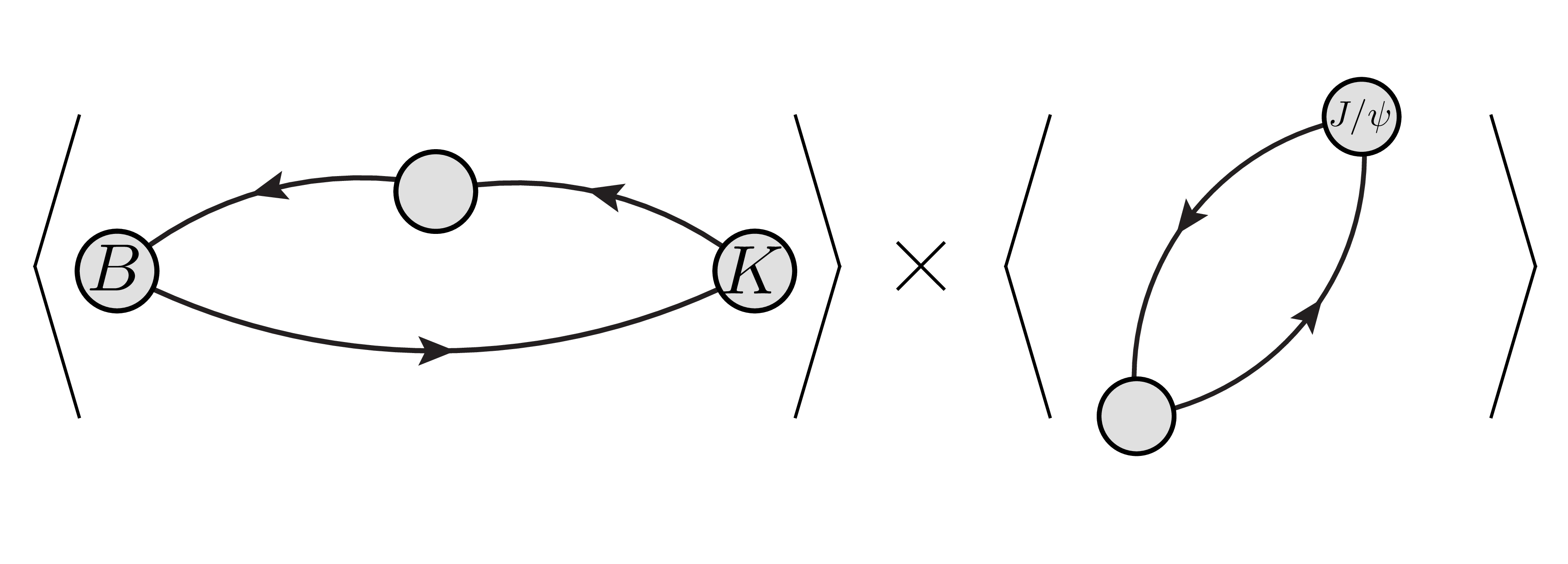}
  \caption{
    Factorization approximation $B\to K \ell^+\ell^-$ with charmonium $J/\psi$ resonances.
}
\label{fig:factori}
\end{center}
\end{figure}

\begin{figure}[tbp]
\begin{center}
  \includegraphics[width=5cm, angle=0]{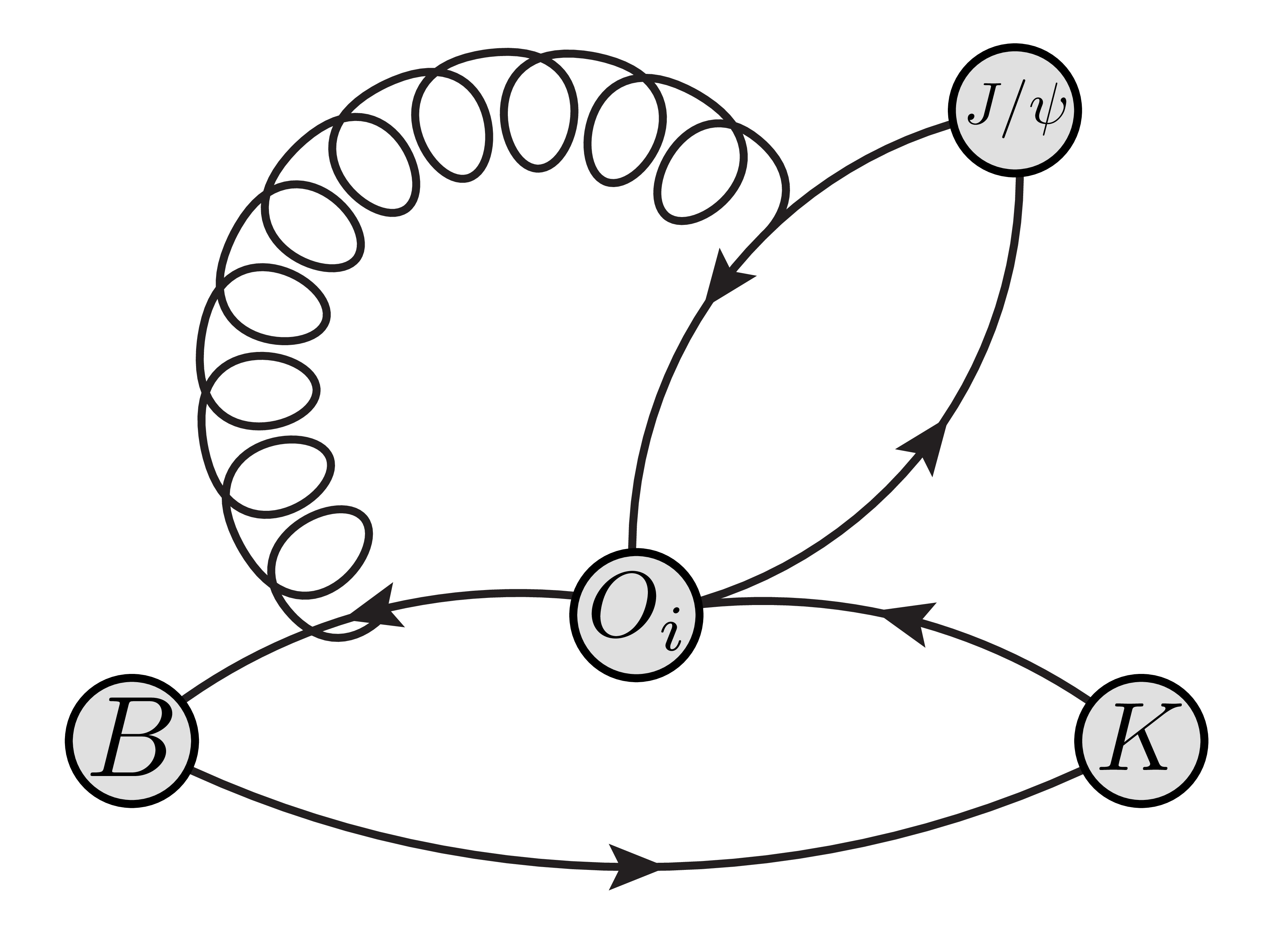}
  \caption{
    Non-factorizable contribution for $B\to K \ell^+\ell^-$.
}
\label{fig:onnceg}
\end{center}
\end{figure}

\begin{figure}[tbp]
\begin{center}
  \includegraphics[width=7cm, angle=0]{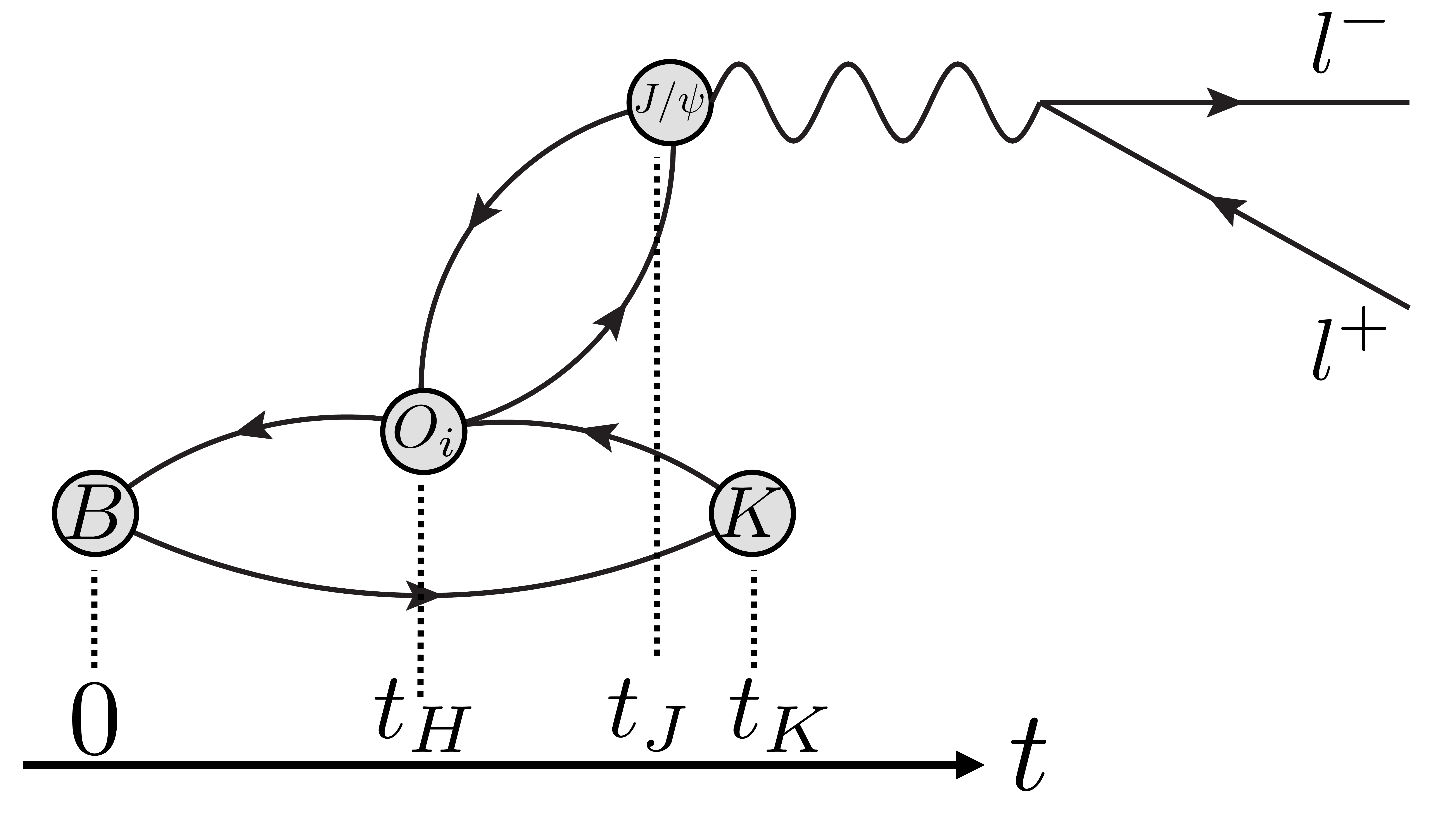}
  \caption{
    Setup of the lattice calculation of the $B\to K \ell^+\ell^-$ amplitude through charmonium $J/\psi$ resonances.
}
\label{fig:BtoKll}
\end{center}
\end{figure}

The factorization approximation assumes that gluon exchanges between the two reduced amplitudes can be ignored.
We define the operators made of color singlet and octet contractions,
$O^{(1)}$ and $O^{(8)}$ respectively, as 
\beqnn{
O^{(1)}
&=&
(\overline{c}_i\gamma_\mu P_- c_i )
(\overline{s}_j\gamma_\mu P_- b_j),\nonumber \\
O^{(8)}
&=&
(\overline{c}_i[T^a]_{ij}\gamma_\mu P_- c_j )
(\overline{s}_k[T^a]_{kl}\gamma_\mu P_- b_l),
}
where generators $T^a$ of $SU(3)$ Lie algebra are introduced with a normalization $[T_a, T_b] = \delta_{ab}/2$.
The factorization assumption corresponds to
\be{
\langle
KJ/\psi |
O^{(1)}
|B\rangle
\simeq
\langle K |
\overline{s}_i\gamma_\mu P_-b_i
|B
\rangle
\langle J/\psi |
\overline{c}\gamma_\mu c
| 0\rangle,
\label{fac1}
}
\be{
\langle
KJ/\psi |
O^{(8)}
|B\rangle
\simeq
0.
\label{fac2}
}
%
Graphically, the factorization can be viewed as in Fig.~\ref{fig:factori}, while the non-factorizable contribution may arise from the diagram shown in Fig.~\ref{fig:onnceg}, for instance.

In order to transform $O^{(1)}$ and $O^{(8)}$ to $O_1 ^c$ and $O_2 ^c$, which appear in the weak effective Hamiltonian, we use the Firtz transformation,
$
\overline{q}_1\gamma_\mu P_- q_2 \overline{q}_3\gamma_\mu P_-q_4
=
\overline{q}_1\gamma_\mu P_- q_4 \overline{q}_3\gamma_\mu P_-q_2.
$
The operator $O^{(1)}$ is reduced to the operator, $O_1 ^c$.
The octet operator $O^{(8)}$ is represented by a linear combination of $O_1 ^c$ and $O_2 ^c$:
\beqnn{
O_1 ^c
&=&
O^{(1)},\nonumber\\
O_2 ^c
&=&
\frac{1}{3}
O^{(1)}
+
2O^{(8)}.
}

In this work, we investigate the validity of the factorization relation (\ref{fac1}) using the lattice calculation.
Namely, we calculate the ratio $R_1$ on the lattice of volume $V$,
\be{
R_1
\equiv
\frac{V\langle K|J_\nu O_1 ^c |B\rangle}
{
\langle 0|
J_\nu J_\mu
|0\rangle
\langle K |
 \overline{s}_j\gamma_\mu P_- b_j
|B\rangle,
}
\label{R1}
}
and see if $R_1\simeq 1$ is a good approximation.
For the other relation (\ref{fac2}), we test the factorization assumption by measuring the ratio between $\langle K|O_2 ^c |B\rangle$ and its factorized form.
%
If the factorization is satisfied for long-range interactions between $B\to K$ decay and charmonium resonances, the ratio $R_{1/3}$ should be $1/3$,
\be{
R_{1/3}
\equiv
\frac{V\langle K|J_\nu O_2 ^c |B\rangle}
{
\langle 0|
J_\nu J_\mu
|0\rangle
\langle K |
 \overline{s}_j\gamma_\mu P_- b_j
|B\rangle
}
\simeq
\frac{1}{3}.
\label{R1/3}
}

\section{Preliminaly result}
%
%
%
%
%
We use the lattice emsemble generated with $N_f = 2 + 1$ dynamical quarks described by the Mobius domain-wall fermion \cite{Brower:2012vk}.
The inverse lattice spacing is $a^{-1} = 3.61$~GeV, and sea quark masses are $am_{ud} = 0.0042$, and $m_s = 0.025$.
In this pilot study, we set the valence light and strange quark mass to $am_{val} = 0.025$.
The charm quark mass is tuned to the physical value $am_c = 0.27287$, and the bottom quark mass is taken slightly lower than the physical value.
This is to eliminate the divergence as we discussed.
With this setup, meson masses are $m_\pi = m_K \simeq 714$~MeV and $m_B\simeq 3.44$~GeV.
%
%
%
%

We set the momenum of the initial $B$ meson state to be $\bm{k} = (0,0,0)$ and the final kaon state at $\bm{p} = (-\frac{2\pi}{L},0,0)$.
Namely, the $c\overline{c}$ system has a momentum $-\bm{p} = (\frac{2\pi}{L},0,0)$.
The energy of two-point correlators for $K$ and $J/\psi$ is $E_K \simeq 854$~MeV, $E_{J/\psi} \simeq 3.13$~GeV.
We set the $B$ meson source at $t=0$, the electromagnetic coupling $J_\mu$ at $t_J = 27$, and $K$ meson source at $t_K=42$.
Here and in the following, we use lattice spacing $a$ as a unit of time, $t$.
The setup is depicted in Fig.~\ref{fig:BtoKll}.
In order to improve the signal, we introduce a parallel shift of the source points on the same configuration.
We use three shifts $t\to t+nT/4$ for each $n = {1,2,3}$.
We prepare $100$ configurations, and then the total number of our statistical data points is $400$.
In this report, our calculation is based on $377$ measurments.
\begin{figure}[tbp]
\begin{center}
  \includegraphics[width=7cm, angle=-90]{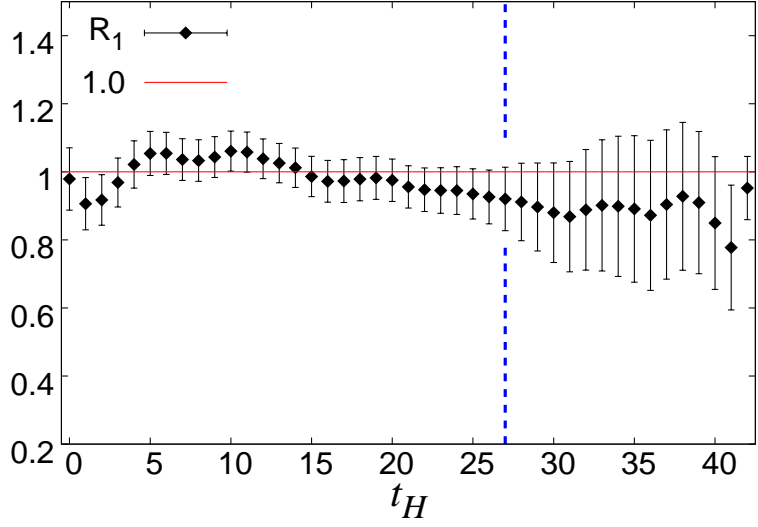}
  \caption{
    The ratio $R_1$ to test the factorization assumption. 
    The electromagnetic current is set at $t_J=27$ as shown by the dashed line.
}
\label{fig:R1}
\end{center}
\end{figure}

Figure~\ref{fig:R1} shows the ratio $R_1$ defined in $(\ref{R1})$.
Our lattice calculation seems to support the factorization assumption within the statistical error of about 10-20\% .
%
\begin{figure}[tbp]
\begin{center}
  \includegraphics[width=7cm, angle=-90]{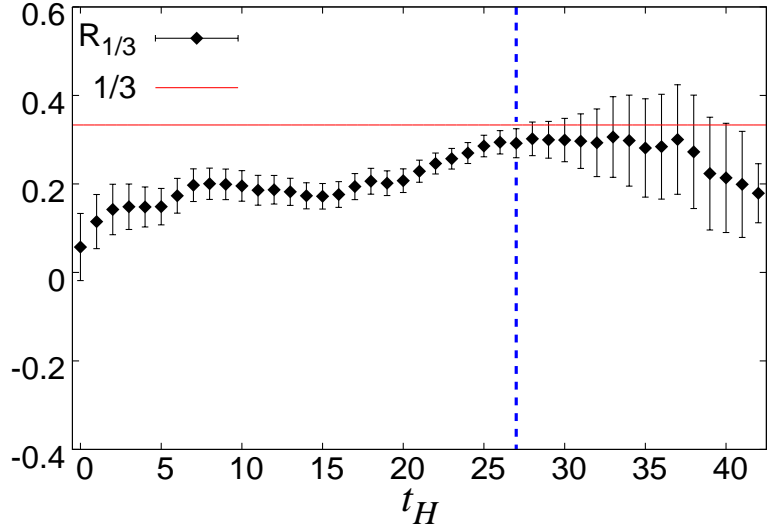}
  \caption{
    The ratio $R_{1/3}$ for testing the factorization method.
    The electromagnetic current is set at $t_J=27$ as shown by the dashed line.
}
\label{fig:R1/3}
\end{center}
\end{figure}
Figure~\ref{fig:R1/3} shows the ratio $R_{1/3}$ defined in $(\ref{R1/3})$.
We find a sizable deviation from 1/3 in the long distance region from the electromagnetic current set at $t = 27$.
It is worth noting that the deviation is found only in the long-distance region.
When $H_\mathrm{eff}$ and $J_\mu$ are close to each other, the relation $R_{1/3} = 1/3$ seems to be satisfied within the statistical error.

\section{Discussions}

To summarize, we calculate the four-point correlator which corresponds to the $B\to K \ell^+\ell^-$ decay through the charmonium resonances $\psi'$s on the lattice.
Our lattice calculation suggests the violation of the factorization in the long-distance regime.
For more conclusive study, we need the renormalization constants of $O_1 ^c$ and $O_2 ^c$ in our lattice calculation.
We should also investigate the case with larger momenta in order to approach the physical decay kinematics.

The $b$ quark mass in this calculation is taken smaller than the physical value in order to satisfy the condition $E_B < E_{J/\psi}+E_K$.
We expect that the amplitude corresponding to the physical $b$ quark mass may be estimated from this setup using the idea of heavy quark effective theory.
Namely, the dynamics of the initial state $B$ meson is largely independent of the $b$ quark mass other than the trivial factor of $\exp{-m_bt}$.


\section*{Acknowledgements}

The lattice QCD simulation has been performed on Blue Gene/Q supercomputer at the High Energy Accelerator Research Organization (KEK) under the Large Scale Simulation Program (Nos. 15/16-09, 16/17-14). 
Oakforest-PACS at JCAHPC under the support of the HPCI System Research Projects.
K. N. is supported by the Grant-in-Aid for JSPS (Japan Society for the Promotion of Science) Research Fellow (No. 18J11457).
This work is supported in part by the Grant-in-Aid of the Japanese Ministry of Education (No. 18H03710).

\bibliographystyle{JHEP.bst}

\bibliography{skeleton_for_arXiv_0121.bbl}


\end{document}